\documentstyle[aps,prd,epsf]{revtex}

\preprint{NUC-MINN-01/...-T}

\newcommand{\be}{\begin{equation}} 
\newcommand{\ee}{\end{equation}}
\newcommand{\ba}{\begin{eqnarray}}
\newcommand{\ea}{\end{eqnarray}}

\title{Magnetic catalysis and anisotropic confinement in QCD}

\author{V.A.~Miransky$^{*}$}

\address{Department of Applied Mathematics, University of Western
Ontario, London, Ontario N6A 5B7, Canada}

\author{I.A.~Shovkovy$^{*}$}

\address{School of Physics and Astronomy, University of Minnesota, 
Minneapolis, MN 55455, USA}

\date{June 19, 2002}

\draft
\begin{document}
\maketitle

\begin{abstract} 
The expressions for dynamical masses of quarks in the chiral limit in QCD
in a strong magnetic field are obtained. A low energy effective action for
the corresponding Nambu-Goldstone bosons is derived and the values of
their decay constants as well as the velocities are calculated. The
existence of a threshold value of the number of colors $N^{thr}_c$,
dividing the theories with essentially different dynamics, is established.
For the number of colors $N_c \ll N^{thr}_c$, an anisotropic dynamics of
confinement with the confinement scale much less than $\Lambda_{QCD}$ and
a rich spectrum of light glueballs is realized.  For $N_c$ of order
$N^{thr}_c$ or larger, a conventional confinement dynamics takes place. It
is found that the threshold value $N^{thr}_c$ grows rapidly with the
magnetic field [$N^{thr}_c \gtrsim 100$ for $|eB| \gtrsim (1\mbox{
GeV})^2$]. In contrast to QCD with a nonzero baryon density, there are no
principal obstacles for examining these results and predictions in lattice
computer simulations. 
\end{abstract}



\section{Introduction}

Since the dynamics of QCD is extremely rich and complicated, it is
important to study this theory under external conditions which provide a
controllable dynamics. On the one hand, this allows one to understand
better the vacuum structure and Green's functions of QCD, and, on the
other hand, there can exist interesting applications of such models in
themselves. The well known examples are hot QCD (for a review see
Ref.~\cite{GPY}) and QCD with a large baryon density (for a review see
Ref. \cite{RW}).

Studies of QCD in external electromagnetic fields had started long ago
\cite{Kawati,SMS}. A particularly interesting case is an external magnetic
field. Using the Nambu-Jona-Lasinio (NJL) model as a low energy effective
theory for QCD, it was shown that a magnetic field enhances the
spontaneous chiral symmetry breakdown. The understanding of this
phenomenon had remained obscure until a universal role of a magnetic 
field as a catalyst of chiral symmetry breaking was established in 
Refs. \cite{GMS1,GMS2}. The general result states that a constant
magnetic field leads to the generation of a fermion dynamical mass (i.e., 
a gap in the one-particle energy spectrum) even at the weakest attractive
interaction between fermions. For this reason, this phenomenon was called
the magnetic catalysis. The essence of the effect is the dimensional
reduction $D\to D-2$ in the dynamics of fermion pairing in a magnetic
field. In the particular case of weak coupling, this dynamics is dominated
by the lowest Landau level (LLL) which is essentially $D-2$ dimensional
\cite{GMS1,GMS2}. The applications of this effect have been considered
both in condensed matter physics \cite{thermo,Khvesh} and cosmology (for
reviews see Ref.~\cite{reviews}).

The phenomenon of the magnetic catalysis was studied in gauge
theories, in particular, in QED \cite{QED1,QED2,Ng,Hong,GS,AFK} 
and in QCD \cite{Sh,Eb,KLW}. In the recent work \cite{KLW}, it has been
suggested that the dynamics underlying the magnetic catalysis in 
QCD is weakly coupled at sufficiently large magnetic fields. In 
this paper, we study this dynamical problem rigorously, from first 
principles. In fact, we show that, at sufficiently strong magnetic 
fields, $|eB| \gg \Lambda_{QCD}^2$, there exists a consistent 
truncation of the Schwinger-Dyson (gap) equation which leads to
a reliable asymptotic expression for the quark mass $m_{q}$. 
Its explicit form reads:
\be
m_{q}^2 \simeq 2 C_{1} |e_{q}B|
\left(c_{q}\alpha_{s}\right)^{2/3}
\exp\left[-\frac{4N_{c}\pi}{\alpha_{s} (N_{c}^{2}-1)
\ln(C_{2}/c_{q}\alpha_{s})}\right],
\label{gap}
\ee
where $e_{q}$ is the electric charge of the $q$-th quark and $N_{c}$
is the number of colors. The numerical factors $C_1$ and $C_2$ equal
$1$ in the leading approximation that we use. Their value, however, can
change beyond this approximation and we can only say that they are 
of order $1$. The constant $c_{q}$ is defined as follows:
\be
c_{q} = \frac{1}{6\pi}(2N_{u}+N_{d})\left|\frac{e}{e_{q}}\right|,
\ee
where $N_{u}$ and $N_{d}$ are the numbers of up and down quark 
flavors, respectively. The total number of quark flavors is $N_{f}
=N_{u}+N_{d}$. The strong coupling $\alpha_{s}$ in the last equation is 
related to the scale $\sqrt{|eB|}$, i.e.,
\be
\frac{1}{\alpha_{s}} \simeq b\ln\frac{|eB|}{\Lambda_{QCD}^2},
\quad \mbox{ where} \quad
b=\frac{11 N_c -2 N_f}{12\pi}. 
\label{coupling}
\ee
We should note that in the leading approximation  
the energy scale $\sqrt{|eB|}$ in Eq. (\ref{coupling}) 
is fixed only up to a factor of order $1$.

As we discuss below, because of the running of $\alpha_{s}$, the value of
the dynamical mass (\ref{gap}) grows very slowly with increasing the value
of the background magnetic field. Moreover, there may exist an intermediate 
region of
fields where the mass {\it decreases} with increasing the magnetic field.
Another, rather unexpected, consequence is that a strong external magnetic
field {\it suppresses} the chiral vacuum fluctuations leading to the
generation of the usual dynamical mass of quarks $m^{(0)}_{dyn} \simeq 300
\mbox{ MeV}$ in QCD without a magnetic field. In fact, in a wide range of
strong magnetic fields $\Lambda^{2} \lesssim B \lesssim (10 \mbox{
TeV})^{2}$ (where $\Lambda$ is the characteristic gap in QCD without the
magnetic field; it can be estimated to be a few times larger than
$\Lambda_{QCD}$), the dynamical mass (\ref{gap}) remains {\it smaller}
than $m^{(0)}_{dyn}$. As it will be shown in Sec. \ref{conf}, this point is
intimately connected with another one: in a strong magnetic field, the
confinement scale, $\lambda_{QCD}$, is much less than the confinement
scale $\Lambda_{QCD}$ in QCD without a magnetic field.

The central dynamical issue underlying this dynamics is the effect of
screening of the gluon interactions in a magnetic field in the region of
momenta relevant for the chiral symmetry breaking dynamics, $m_{q}^2 \ll
|k^2| \ll |eB|$. In this region, gluons acquire a mass $M_{g}$ of order
$\sqrt{N_{f}\alpha_{s}|e_{q}B|}$. This allows to separate the dynamics 
of the magnetic catalysis from that of confinement. More rigorously, 
$M_{g}$ is the mass of a quark-antiquark composite state coupled to
the gluon field. The appearance of such mass resembles pseudo-Higgs effect
in the $1+1$ dimensional {\it massive} QED (massive Schwinger model)
\cite{Sch} (see below).

Since the background magnetic field breaks explicitly the global chiral
symmetry that interchanges the up and down quark flavors, the chiral
symmetry in this problem is $SU(N_{u})_{L}\times SU(N_{u})_{R} \times
SU(N_{d})_{L}\times SU(N_{d})_{R}\times U^{(-)}(1)_{A}$.
The $U^{(-)}(1)_{A}$ is connected with the current which is an
anomaly free linear combination 
of the $U^{(d)}(1)_{A}$ and $U^{(u)}(1)_{A}$ currents.
[The $U^{(-)}(1)_{A}$ symmetry is of course absent if either
$N_d$ or $N_u$ is equal to zero].
The generation of quark masses
breaks this symmetry spontaneously down to $SU(N_{u})_{V}\times
SU(N_{d})_{V}$ and, as a result, $N_{u}^{2}+N_{d}^{2}-1$ gapless
Nambu-Goldstone (NG) bosons occur. In Sec. \ref{NG}, we derive the
effective action for the NG bosons and calculate their decay constants and
velocities.

The present analysis is heavily based on the analysis of the magnetic
catalysis in QED done by Gusynin, Miransky, and Shovkovy \cite{QED2}.  
A crucial difference is of course the property of asymptotic
freedom and confinement in QCD. In connection with that, our second major
result is the derivation of the low energy effective action for gluons in
QCD in a strong magnetic field [see Eq.~(\ref{gluon-action}) below]. The
characteristic feature of this action is its anisotropic dynamics. In
particular, the strength of static (Coulomb like) forces along the
direction parallel to the magnetic field is much larger than that in the
transverse directions.  Also, the confinement scale in this theory is much
less than that in QCD without a magnetic field. This features imply a rich
and unusual spectrum of light glueballs in this theory.

A special and interesting case is QCD with a large number of colors, in
particular, with $N_c \to \infty$ (the 't Hooft limit). In this limit, the
mass of gluons goes to zero and the expression for the quark mass becomes
essentially different [see Eq. (\ref{m_q}) in Sec. \ref{infty}]. In fact,
it will be shown that, for any value of an external magnetic field, there
exists a threshold value $N^{thr}_{c}$, rapidly growing with $|eB|$ 
[e.g., $N^{thr}_c \gtrsim 100$ for $|eB| \gtrsim (1\mbox{ GeV})^2$].
For $N_c$ of the order
$N^{thr}_{c}$ or larger, the gluon mass becomes small and irrelevant for
the dynamics of the generation of a quark mass. As a result, expression
(\ref{m_q}) for $m_q$ takes place for such large $N_c$. The confinement
scale in this case is close to $\Lambda_{QCD}$. Still, as is shown in
Sec. \ref{infty}, the dynamics of chiral symmetry breaking is under 
control in this limit if the magnetic field is sufficiently strong.

It is important that, unlike the case of QCD with a nonzero baryon
density, there are no principal obstacles for checking all these results
and predictions in lattice computer simulations of QCD in a magnetic
field.

\section{Magnetic catalysis in QCD}
\label{mag-cat}

We begin by considering the Schwinger-Dyson (gap) equation for 
the quark propagator. It has the following form:
\ba
G^{-1}(x,y) &=& S^{-1}(x,y) + 4\pi\alpha_{s} \gamma^{\mu}\int G(x,z)
\Gamma^{\nu}(z,y,z^{\prime}) {\cal D}_{\nu\mu}(z^{\prime},x) 
d^{4} z d^{4} z^{\prime}, 
\label{SD}     
\ea
where $S(x,y)$ and $G(x,y)$ are the bare and full fermion propagators
in an external magnetic field, ${\cal D}_{\nu\mu}(x,y) $ is the 
full gluon propagator and $\Gamma^{\nu}(x,y,z)$ is the full amputated
vertex function. Since the coupling $\alpha_s$ related to the scale 
$|eB|$ is small, one might think that the rainbow (ladder) approximation
is reliable in this problem. However, this is not the case.
Because of the (1+1)-dimensional form of the fermion propagator
in the LLL approximation, there are relevant higher order 
contributions \cite{QED1,QED2}. Fortunately one can solve this problem.
First of all, an important feature of the quark-antiquark pairing dynamics 
in QCD in a strong magnetic field is that this dynamics is essentially 
abelian. This feature is provided by the form of the polarization 
operator of gluons in this theory. The point is that the dynamics 
of the quark-antiquark pairing is mainly induced 
in the region of momenta $k$ 
much less than $\sqrt{|eB|}$. This implies that the magnetic field 
yields a dynamical ultraviolet cutoff in this problem. On the other
hand, while the contribution of (electrically neutral) gluons
and ghosts in the polarization operator is proportional to
$k^2$, the fermion contribution is proportional to $|e_{q}B|$
\cite{QED2}. As a result, the fermion contribution dominates
in the relevant region with $k^2 \ll |eB|$. 

This observation implies that there are three, dynamically very 
different, scale regions in this problem. The first one is the region 
with the energy scale above the magnetic scale $\sqrt{|eB|}$.
In that region, the dynamics is essentially the same as in QCD
without a magnetic field. In particular, the running coupling 
decreases logarithmically with increasing the energy scale there. The
second region is that with the energy scale below the magnetic scale
but much larger than the dynamical mass $m_{q}$. In this region,
the dynamics is abelian like and, therefore, the dynamics of the 
magnetic catalysis is similar to that in QED in a magnetic field. 
At last, the third region is the region with the energy scale less 
than the gap. In this region, quarks decouple and a confinement 
dynamics for gluons is realized.
 
Let us first consider the intermediate region relevant for the 
magnetic catalysis. As was indicated above, the important ingredient 
of this dynamics is a large contribution of fermions to the 
polarization operator. It is large because of an (essentially)
1+1 dimensional form of the fermion propagator in a strong magnetic
field. Its explicit form can be obtained by modifying
appropriately the expression for the polarization operator
in QED in a magnetic field \cite{QED2}:
\begin{eqnarray}
{\cal P}^{AB,\mu\nu} \simeq \frac{\alpha_{s}}{6\pi}
\delta^{AB} \left(k_{\parallel}^{\mu}
k_{\parallel}^{\nu}-k_{\parallel}^{2}g_{\parallel}^{\mu\nu}\right)
\sum_{q=1}^{N_{f}}\frac{|e_{q}B|}{m^{2}_{q}},
\quad \mbox{for} \quad |k_{\parallel}^2| \ll m_{q}^2,
\label{Pi-IR}\\
{\cal P}^{AB,\mu\nu} \simeq -\frac{\alpha_{s}}{\pi}
\delta^{AB} \left(k_{\parallel}^{\mu}
k_{\parallel}^{\nu}-k_{\parallel}^{2}g_{\parallel}^{\mu\nu}\right)
\sum_{q=1}^{N_{f}}\frac{|e_{q}B|}{{k_{\parallel}^2}}, 
\quad \mbox{for} \quad m_{q}^2 \ll |k_{\parallel}^2|\ll |eB|,
\label{Pi-UV}
\end{eqnarray}
where $g_{\parallel}^{\mu\nu}\equiv \mbox{diag}(1,0,0,-1)$ is
the projector onto the longitudinal subspace,
and $k_{\parallel}^{\mu}\equiv g_{\parallel}^{\mu\nu} k_{\nu}$
(the magnetic field is in the $x^3$ direction).
Similarly, we introduce the orthogonal projector $g_{\perp}^{\mu\nu}\equiv
g^{\mu\nu}-g_{\parallel}^{\mu\nu}=\mbox{diag}(0,-1,-1,0)$ and
$k_{\perp}^{\mu}\equiv g_{\perp}^{\mu\nu} k_{\nu}$ that
we shall use below. Notice that quarks in a strong magnetic field
do not couple to the transverse subspace spanned by $g_{\perp}^{\mu\nu}$
and $k_{\perp}^{\mu}$. This is because in a strong magnetic field 
only the quark from the LLL matter and they couple only to the
longitudinal components of the gluon field. The latter property 
follows from the fact that spins of the LLL quarks are polarized 
along the magnetic field \cite{QED1}.  

The expressions (\ref{Pi-IR}) and (\ref{Pi-UV}) coincide with 
those for the polarization operator in the massive Schwinger
model if the parameter $\alpha_{s} |e_{q}B|/2$ here is replaced by 
the dimensional coupling $\alpha_{1}$ of $QED_{1+1}$. As in the 
Schwinger model, Eq.~(\ref{Pi-UV}) implies that there is a massive 
resonance in the $k_{\parallel}^{\mu}k_{\parallel}^{\nu}
-k_{\parallel}^{2} g_{\parallel}^{\mu\nu}$ component of the gluon 
propagator. Its mass is
\be
M_{g}^2= \sum_{q=1}^{N_{f}}\frac{\alpha_{s}}{\pi}|e_{q}B|=
(2N_{u}+N_{d}) \frac{\alpha_{s}}{3\pi}|eB|.
\label{M_g}
\ee
This is reminiscent of the pseudo-Higgs effect in the 
(1+1)-dimensional massive QED. It is not the genuine Higgs effect 
because there is no complete screening of the color charge in the 
infrared region with $|k_{\parallel}^2|\ll m_{q}^2$. This can 
be seen clearly from Eq.~(\ref{Pi-IR}). Nevertheless, the pseudo-Higgs 
effect is manifested in creating a massive resonance and this 
resonance provides the dominant forces leading to chiral
symmetry breaking.  

Now, after the abelian like structure of the dynamics in this 
problem is established, we can use the results of the analysis in 
QED in a magnetic field \cite{QED2} by introducing appropriate 
modifications. The main points of the analysis are: (i) the so
called improved rainbow approximation is reliable in this problem
provided a special non-local gauge is used in the analysis, 
and (ii) for a small coupling $\alpha_{s}$ 
($\alpha$ in QED), the relevant region of momenta in
this problem is $m_{q}^2 \ll |k^2| \ll |eB|$. We recall 
that in the improved rainbow approximation the vertex 
$\Gamma^{\nu}(x,y,z)$ is taken to be bare and the gluon propagator 
is taken in the one-loop approximation. Moreover, as we argued 
above, in this intermediate region of momenta, only the contribution
of quarks to the gluon polarization tensor (\ref{Pi-UV}) matters.
[It is appropriate to call this approximation the 
``strong-magnetic-field-loop improved rainbow approximation". It
is an analog of the hard-dense-loop improved rainbow approximation in
QCD with a nonzero baryon density \cite{hard}].
As to the modifications, they are purely kinematic: the overall 
coupling constant in the gap equation $\alpha$ and the dimensionless
combination $M_{\gamma}^2/|eB|$ in QED have to be replaced by 
$\alpha_s(N_{c}^2-1)/2N_{c}$ and $M_{g}^2/|e_{q} B|$, respectively. 
This leads us to the expression (\ref{gap}) for the dynamical gap.
   
After expressing the magnetic field in terms of the running coupling, 
the result for the dynamical mass takes the following convenient form:
\be
m_{q}^2 \simeq 2C_{1}\left|\frac{e_{q}}{e}\right| \Lambda_{QCD}^2 
\left(c_{q}\alpha_{s}\right)^{2/3}
\exp\left[\frac{1}{b\alpha_{s}}-\frac{4N_{c}\pi}{\alpha_{s} 
(N_{c}^{2}-1) \ln(C_{2}/c_{q}\alpha_{s})}\right].
\label{gap-vs-alpha}
\ee
As is easy to check, the dynamical mass of the $u$-quark 
is considerably larger than that of the $d$-quark. It is also
noticeable that the values of the $u$-quark dynamical mass
becomes comparable to the vacuum value 
$m^{(0)}_{dyn}\simeq 300 \mbox{ MeV}$
only when the coupling constant gets as small as $0.05$. 

\begin{figure}
\begin{center}
\epsfxsize=8.0cm
\epsffile[88 4 488 252]{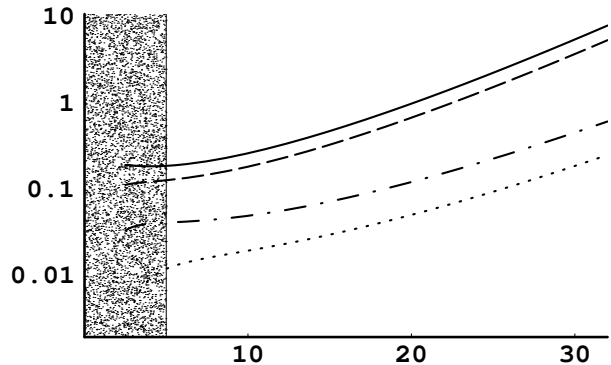}
\caption{The dynamical masses of quarks as functions of 
$\ln(|eB|/\Lambda_{QCD}^2)$ for $N_{c}=3$ and two different
values of $N_{f}=N_{u}+N_{d}$: (i) masses of $u$-quark (solid line) 
and $d$-quark (dash-dotted line) for $N_{u}=1$ and $N_{d}=2$;
(ii) masses of $u$-quark (dashed line) and $d$-quark (dotted line) 
for $N_{u}=2$ and $N_{d}=2$. The result may not be reliable in the 
weak magnetic field region (shaded) where some of the approximations 
break. The values of masses are given in units 
of $\Lambda_{QCD} = 250 \mbox{ MeV}$.}
\label{fig:rho_vs_LogB}
\end{center}
\end{figure}

Now, by trading the coupling constant for the magnetic field scale
$|eB|$, we get the dependence of the dynamical mass 
on the value of the external field. The numerical results 
are presented in  Fig. \ref{fig:rho_vs_LogB} [we used 
$C_{1} = C_{2} = 1$ in Eq. (\ref{gap-vs-alpha})]. 
As one can see in Fig. \ref{fig:rho_vs_LogB}, the value of the quark gap
in a wide window of strong magnetic fields, $\Lambda_{QCD}^{2}\ll |eB|
\lesssim (10 \mbox{ TeV})^{2}$, remains smaller than the dynamical mass of
quarks $m^{(0)}_{dyn} \simeq 300 \mbox{ MeV}$ in QCD without a magnetic field.  
In other words, the chiral condensate is partially {\it suppressed} for
those values of a magnetic field.
The explanation of this, rather unexpected, result is
actually simple. The magnetic field leads to the mass $M_g$ (\ref{M_g})  
for gluons. In a strong enough magnetic field,
this mass becomes larger than the characteristic gap
$\Lambda$ in QCD without a magnetic field ($\Lambda$, playing the role of
a gluon mass, can be estimated as a few times larger than
$\Lambda_{QCD}$). This, along with the property of the asymptotic freedom
(i.e., the fact that $\alpha_{s}$ decreases with increasing the magnetic
field), leads to the suppression of the chiral condensate \cite{footnote}.

This point also explains why our result for the gap is so different from
that in the NJL model in a magnetic field \cite{Kawati}. Recall that, in
the NJL model, the gap logarithmically (i.e., much faster than in the
present case) grows with a magnetic field. This is the related to the
assumption that both the dimensional coupling constant $G =
g/\Lambda^2$ (with $\Lambda$ playing a role similar to that of the gluon
mass in QCD), as well as the scale $\Lambda$ do not dependent on the value
of the magnetic field. Therefore, in that model, in a strong enough
magnetic field, the value of the chiral condensate is overestimated.

The picture which emerges from this discussion is the following.
For values of a magnetic field $|eB| \alt \Lambda^2$ the
dynamics in QCD should be qualitatively similar to that in
the NJL model. For strong values of the field, however, it
is essentially different, as was described above. This in turn
suggests that there should exist an intermediate region of
fields where the dynamical masses of quarks decreases with
increasing the background magnetic field.

\section{Effective action of NG bosons}
\label{NG}
   
The presence of the background magnetic field breaks explicitly the global
chiral symmetry that interchanges the up and down quark flavors. This is
related to the fact that the electric charges of the two sets of quarks
are different. However, the magnetic field does not break the global
chiral symmetry of the action completely.  In particular, in the model
with the $N_{u}$ up quark flavors and the $N_{d}$ 
down quark flavors, the action is invariant under the chiral symmetry
$SU(N_{u})_{L}\times SU(N_{u})_{R} \times SU(N_{d})_{L}\times
SU(N_{d})_{R}\times U^{(-)}(1)_{A}$. 
The $U^{(-)}(1)_{A}$ is connected with the current which is an
anomaly free linear combinaton
of the $U^{(d)}(1)_{A}$ and $U^{(u)}(1)_{A}$ currents.
[The $U^{(-)}(1)_{A}$ symmetry is of course absent if either
$N_d$ or $N_u$ is equal to zero].

The global chiral symmetry of the action is broken spontaneously down to
the diagonal subgroup $SU(N_{u})_{V}\times SU(N_{d})_{V}$ when 
dynamical masses of quarks are generated. In agreement with the
Goldstone theorem, this leads to the
appearance of $N_{u}^{2}+N_{d}^{2}-1$ number of the NG gapless excitations
in the low-energy spectrum of QCD in a strong magnetic field. Notice
that there is also a pseudo-NG boson connected with the conventional
(anomalous) $U(1)_A$ symmetry which can be rather light in a
sufficiently strong magnetic field.
   
Now, in the chiral limit, the general structure of the low energy action
for the NG bosons could be easily established from the symmetry arguments
alone. First of all, such an action should be invariant with respect to
the space-time symmetry $SO(1,1)\times SO(2)$ which is left unbroken by
the background magnetic field [here the SO(1,1) and the SO(2) are
connected with Lorentz boosts in the $x_{0}-x_{3}$ hyperplane and
rotations in the $x_{1}-x_{2}$ plane, respectively]. Besides that, the
low-energy
action should respect the original chiral symmetry $SU(N_{u})_{L}\times
SU(N_{u})_{R} \times SU(N_{d})_{L}\times SU(N_{d})_{R}\times
U^{(-)}(1)_{A}$. 
These
requirements lead to the following general form of the action:
\ba
{\cal L}_{NG} &\simeq &\frac{f_{u}^{2}}{4}
\mbox{tr} \left( g_{\parallel}^{\mu\nu}
\partial_{\mu}\Sigma_{u}\partial_{\nu}\Sigma_{u}^{\dagger}
+v_{u}^{2} g_{\perp}^{\mu\nu}
\partial_{\mu}\Sigma_{u}\partial_{\nu}\Sigma_{u}^{\dagger}\right)
\nonumber\\
&+&\frac{f_{d}^{2}}{4}
\mbox{tr} \left(g_{\parallel}^{\mu\nu}
\partial_{\mu}\Sigma_{d}\partial_{\nu}\Sigma_{d}^{\dagger}
+v_{d}^{2} g_{\perp}^{\mu\nu}
\partial_{\mu}\Sigma_{d}\partial_{\nu}\Sigma_{d}^{\dagger}\right)
\nonumber\\
&+& \frac{\tilde{f}^{2}}{4}
\left(g_{\parallel}^{\mu\nu}
\partial_{\mu}\tilde{\Sigma}\partial_{\nu}
\tilde{\Sigma}^{\dagger}
+\tilde{v}^{2} g_{\perp}^{\mu\nu}
\partial_{\mu}\tilde{\Sigma}\partial_{\nu}\tilde{\Sigma}^{\dagger}\right).
\label{low-e-NG}
\ea
The unitary matrix fields $\Sigma_{u}\equiv \exp \left(
i\sum_{A=1}^{N_{u}^2-1}\lambda^{A}\pi_{u}^{A}/f_{u} \right)$,
$\Sigma_{d}\equiv \exp \left( i\sum_{A=1}^{N_{d}^2-1}
\lambda^{A}\pi_{d}^{A}/f_{d} \right)$, and 
$\tilde{\Sigma} \equiv \exp 
\left({i\sqrt{2}}\tilde{\pi}/\tilde{f} \right)$ 
describe the NG bosons in the up,
down, and $U^{(-)}(1)_{A}$
sectors of the original theory. The decay constants 
$f_{u}, f_{d}, \tilde{f}$ and transverse velocities $v_{u}, v_{d}, 
\tilde{v}$
can be calculated by using the standard field theory formalism (for a
review, see for example the book \cite{M}). Let us first consider 
the $N_{u}^{2}+N_{d}^{2}-2$ NG bosons in the up and down sectors,
assigned to the adjoint representation of 
the $SU(N_{u})_{V}\times SU(N_{d})_{V}$ symmetry. 
The basic relation is
\be
\delta^{AB}P_{q}^{\mu} f_{q} = -i\int 
\frac{d^{4}k}{(2\pi)^{4}}  \mbox{tr}
\left[\gamma^{\mu}\gamma^{5}\frac{\lambda^A}{2}
\chi^{B}_{q}(k,P)\right],
\label{decay} 
\ee
where $P_{q}^{\mu}= \left(P^{0}, v_{q}^{2}\vec{P_{\perp}}, P^{3}\right)$
and $\chi^{A}_{q}(k,P)$ is the Bethe-Salpeter (BS) wave function of the NG
bosons ($P$ is the momentum of their center of mass). In the weakly
coupled dynamics at hand, one could use an analogue of the Pagels-Stokar 
approximation \cite{PS,M}. 
In this approximation, the BS wave function is
determined from the Ward identities for axial currents. In
fact, the calculation of the decay constants and velocities of NG bosons 
resembles closely the calculation in the case of a color superconducting 
dense quark matter \cite{MSW}. In the LLL approximation, the final result
in Euclidean space is
\ba
f_{q}^2 &=&  4N_c\int \frac{d^{2}k_{\perp} 
d^{2}k_{\parallel}}{(2\pi)^4}
\exp\left(-\frac{k^{2}_{\perp}}{|e_{q}B|}\right)
\frac{m_{q}^2}{(k^{2}_{\parallel} + m_{q}^2)^2},
\quad v_{q}=0.
\label{Pagels}
\ea
The evaluation of this integral is straightforward. As a result,
we get 
\ba
f_{u}^{2} &=& \frac{N_c}{6\pi^{2}}|eB|,\\ 
f_{d}^{2} &=& \frac{N_c}{12\pi^{2}}|eB|.
\ea
The remarkable fact is that the decay constants are nonzero
even in the limit when the dynamical masses of quarks 
approach zero. The reason of that is the $1+1$ dimensional
character of this dynamics: as one can see from expression
(\ref{Pagels}), in the limit $m_{q} \to 0$, the infrared
singularity in the integral cancels the mass $m_q$ in the
numerator. A similar situation takes place in color
superconductivity \cite{SS,MSW}: in that case the
$1+1$ dimensional character of the dynamics is provided by
the Fermi surface. 

Notice that the transverse velocities of the NG bosons are equal to zero.
This is also a consequence of the $1+1$ dimensional structure of the quark
propagator in the LLL approximation.  The point is that quarks can move in
the transverse directions only by hopping to higher Landau levels. Taking
into account higher Landau levels would lead to nonzero velocities
suppressed by powers of $|m_{q}|^{2}/|eB|$. In fact, the explicit form of
the velocities was derived in the weakly coupled NJL model in an
external magnetic field [see Eq.~(65) in the second paper of 
Ref.~\cite{QED1}].
It is
\be
v_{u,d}^{2} \sim \frac{|m_{u,d}|^{2}}{|eB|}
\ln\frac{|eB|}{|m_{u,d}|^{2}} \ll 1.
\ee
A similar expression should take place also for the transverse 
velocities of the NG bosons in QCD.

Now, let us turn to the NG boson connected with the spontaneous
breakdown of the $U^{(-)}(1)_{A}$. It is a 
$SU(N_{u})_{V}\times SU(N_{d})_{V}$ singlet. Neglecting the
anomaly, we would actually get two NG singlets, connected
with the up and down sectors, respectively. Their decay constants and
velocities would be given by expression (\ref{decay}) in
which $\lambda^A$ has to be replaced by $\lambda^0$.
The latter is proportional to the unit matrix and normalized 
as the $\lambda^A$ matrices:
$\mbox{tr}[(\lambda^0)^2]=2$. It is clear that their
decay constants and velocities would be the same as for the
NG bosons from the adjoint representation. Taking now into
account the anomaly, we find that the anomaly free  $U^{(-)}(1)_{A}$
current is connected with the traceless matrix
$\tilde{\lambda}^{0}/2 \equiv (\sqrt{N_{d}/N_{f}}\lambda^{0}_{u} -
\sqrt{N_{u}/N_{f}}\lambda^{0}_{d})/2$. Therefore the genuine NG
singlet $|1 \rangle$ is expressed through 
those two singlets, $|1,d \rangle$ and
$|1,u \rangle$, as 
$|1 \rangle = \sqrt{N_{d}/N_{f}}|1,u \rangle - 
\sqrt{N_{u}/N_{f}}|1,d \rangle$. This
implies
that
its decay constant is
\be 
\tilde{f}^{2} = \frac{(N_{d} f_{u} + 
N_{u} f_{d})^2}{N^{2}_{f}}=
\frac{(\sqrt{2}N_{d} + N_{u})^{2}N_c}{12\pi^{2}N^{2}_{f}}|eB|.
\ee
Its transverse velocity is of course zero in the LLL approximation.

\section{Anisotropic confinement of gluons}
\label{conf}

Let us now turn to the infrared region with $|k|\lesssim m_d$, where all
quarks decouple (notice that we take here the smaller mass of $d$
quarks). In that region, a pure gluodynamics realizes. However, its
dynamics is quite unusual. The point is that although gluons are
electrically neutral, their dynamics is strongly influenced by an external
magnetic field, as one can see from expression (\ref{Pi-IR}) for their
polarization operator. In a more formal language, while quarks decouple
and do not contribute into the equations of the renormalization group in
that infrared region, their dynamics strongly influence the boundary
(matching) conditions for those equations at $k \sim m_d$.

A conventional way to describe this dynamics is the method of the low
energy effective action. By taking into account the polarization effects
due to the background magnetic field, we arrive at the following quadratic
part of the low-energy effective action of gluons:
\be
{\cal L}^{(2)}_{g,eff} = -\frac{1}{2} \sum_{A=1}^{N_{c}^2-1}
A^{A}_{\mu}(-k) \left[g^{\mu\nu}k^2-k^{\mu}k^{\nu}+\kappa
\left(g_{\parallel}^{\mu\nu}k_{\parallel}^2-k_{\parallel}^{\mu}
k_{\parallel}^{\nu}\right)\right] A^{A}_{\nu}(k),
\label{L-eff-2}
\ee
where
\be
\kappa=\frac{\alpha_{s}}{6\pi}
\sum_{q=1}^{N_{f}}\frac{|e_{q}B|}{m^{2}_{q}}=
\frac{1}{12C_{1}\pi} \sum_{q=1}^{N_{f}}
\left(\frac{\alpha_{s}}{c_{q}^{2}}\right)^{1/3}
\exp\left(\frac{4N_{c}\pi}{\alpha_{s} (N_{c}^{2}-1)
\ln(C_{2}/c_{q}\alpha_{s})}\right) \gg 1.
\ee
By making use of the quadratic part of the action as well
as the requirement of the gauge invariance, we could easily
restore the whole low-energy effective action (including
self-interactions) as follows:
\be
{\cal L}_{gl} \simeq  \frac{1}{2} \sum_{A=1}^{N_{c}^2-1}
\left(\vec{E}_{\perp}^{A} \cdot \vec{E}_{\perp}^{A}
+\epsilon E_{3}^{A} E_{3}^{A}
- \vec{B}_{\perp}^{A} \cdot \vec{B}_{\perp}^{A}
-B_{3}^{A} B_{3}^{A}\right),
\label{gluon-action}
\ee
where the (chromo-) dielectric constant $\epsilon \equiv 1+\kappa$
was introduced. Also, we introduced the notation for the
chromo-electric and chromo-magnetic fields as follows:
\ba
E_{i}^{A} &=& \partial_{0} A_{i}^{A} - \partial_{i} A_{0}^{A}
+g f^{ABC} A_{0}^{B} A_{i}^{C}, \\
B_{i}^{A} &=& \frac{1}{2}\varepsilon_{ijk}
\left(\partial_{j} A_{k}^{A} - \partial_{k} A_{j}^{A}
+g f^{ABC} A_{j}^{B} A_{k}^{C}\right).
\ea
This low energy effective action is relevant for momenta $|k| \lesssim
m_d$. Notice the following important feature of the action: the coupling
$g$, playing here the role of the ``bare" coupling constant related to the
scale $m_d$, coincides with the value of the vacuum QCD coupling related
to the scale $\sqrt{|eB|}$ (and {\it not} to the scale $m_d$). This is
because
$g$ is determined from the matching condition at $|k| \sim m_d$, the lower
border of the intermediate region $m_d \lesssim |k| \lesssim \sqrt{|eB|}$,
where,
because of the pseudo-Higgs effect, the running of the coupling is
essentially frozen. Therefore the ``bare" coupling $g$ indeed coincides
with the value of the vacuum QCD coupling related to 
the scale $\sqrt{|eB|}$: $g = g_s$. Since this value is much less that
that of the vacuum QCD coupling
related to the scale $m_d$, this implies that the confinement scale
$\lambda_{QCD}$ of the action (\ref{gluon-action}) should be much less
than $\Lambda_{QCD}$ in QCD without a magnetic field.

Actually, this consideration somewhat simplifies the real
situation. Since the
LLL quarks couple to the longitudinal components of the
polarization operator, only the effective coupling connected
with longitudinal gluons is frozen. For transverse gluons, there
should be a logarithmic running
of their effective coupling. It is clear, however, that this
running should be quite different from that in the vacuum QCD.
The point is that the time like gluons are now massive and 
their contribution in the running in the intermediate region
is severely reduced. On the other hand, 
because of their negative norm, just the time like gluons
are the major players in producing the antiscreening running
in QCD (at least in covariant gauges). 
Since now they effectively decouple,
the running of the effective coupling for the transverse gluons
should slow down. It is even not inconceivable that the
antiscreening running can be transformed into a screening one.  
In any case, one should expect that the value of the transverse
coupling related to the matching scale $m_d$ will be also essentially
reduced in comparison with that in the vacuum QCD. Since the
consideration in this section is rather qualitative, 
we adopt  the simplest scenario with the value of the transverse coupling
at the matching scale $m_d$ also
coinciding with $g_s$.  

In order to determine the new confinement scale $\lambda_{QCD}$, one
should consider the contribution of gluon loops in the perturbative loop
expansion connected with the {\it anisotropic} action
(\ref{gluon-action}), a hard problem being outside the scope of this
paper. Here we will get an estimate of $\lambda_{QCD}$, without 
studying the loop expansion in detail.
Let us start from calculating the interaction potential
between two static quarks in this theory.
It reads:
\be
V(x,y,z) \simeq \frac{g_{s}^{2}}
{4\pi\sqrt{z^2+\epsilon (x^{2}+y^{2})}}.
\label{potent}
\ee
Because of the dielectric constant, this Coulomb like interaction is
anisotropic in space: it is suppressed by a factor of $\sqrt{\epsilon}$ in
the transverse directions compared to the interaction in the direction of
the magnetic field. The potential (\ref{potent}) corresponds to the
classical, tree, approximation which is good only in the region of
distancies much smaller than the confinement radius $r_{QCD} \sim
\lambda^{-1}_{QCD}$. Deviations from this interaction are described by
loop corrections. Let us estimate the value of a fine structure constant
connected with the perturbative loop expansion.

First of all, because of the form of the potential (\ref{potent}), the
effective coupling constants connected with the parallel and transverse
directions are different: while the former is equal to
$g^{\parallel}_{eff} = g_s$, the latter is $g^{\perp}_{eff} =
g_s/\epsilon^{1/4}$. On the other hand, the loop expansion parameter (fine
structure constant) is $g_{eff}^2/4\pi v_g$, where $v_g$ is the velocity
of gluon quanta. Now, as one can notice from Eq.~(\ref{L-eff-2}), while
the velocity of gluons in the parallel direction is equal to the velocity
of light $c = 1$, there are gluon quanta with the velocity $v_{g}^{\perp}
= 1/\sqrt{\epsilon}$ in the transverse directions. This seems to suggest
that the fine structure coupling may remain the same, or nearly the same,
despite the anisotropy: the factor $\sqrt{\epsilon}$ in
$(g^{\perp}_{eff})^2$ will be cancelled by the same factor in
$v_{g}^{\perp}$. Therefore the fine structure constant can be estimated as
$\alpha_{s} = g^{2}_{s}/4\pi$ (although, as follows from
Eq.~(\ref{L-eff-2}), there are quanta with the velocity $v_{g}^{\perp} =
1$, their contribution in the perturbative expansion is suppressed by the
factor $1/\sqrt{\epsilon}$).

This consideration is of course far from being quantitative. Introducing
the magnetic field breaks the Lorentz group $SO(3,1)$ down to $SO(1,1)
\times SO(2)$, and it should be somehow manifested in the perturbative
expansion. Still, we believe, this consideration suggests that the
structure of the perturbative expansion in this theory can be
qualitatively similar to that in the vacuum QCD, modulo the important
variation: while in the vacuum QCD $\alpha_s$ is related to the scale
$|eB|$, it is now related to much smaller scale $m_d$.

By making use of this observation,
we will approximate the running in   
the low-energy region by a vacuum-like running:
\be
\frac{1}{\alpha_{s}^{\prime}(\mu)} = \frac{1}{\alpha_{s}}
+b_{0} \ln\frac{\mu^{2}}{m_{d}^{2}}, \quad \mbox{ where} \quad
b_{0} =\frac{11 N_{c}}{12\pi},
\ee
where the following condition was imposed: 
$\alpha_{s}^{\prime}(m_d)=\alpha_{s}$. From this running law, we estimate
the new confinement scale,
\be
\lambda_{QCD} \simeq m_d
\left(\frac{\Lambda_{QCD}}{\sqrt{|eB|}}\right)^{b/b_{0}}.
\label{lambda}
\ee
We emphasize again that expression (\ref{lambda}) is just an
estimate of the new confinement scale. In particular, both
the exponent, taken here to be equal to $b/b_{0}$, and the
overall factor in this expression, taken here to be equal $1$,
should be considered as being fixed only up to a factor of order one.

The hierarchy $\lambda_{QCD} \ll \Lambda_{QCD}$ is intimately connected
with a somewhat puzzling point that the pairing dynamics decouples
from the confinement dynamics
despite it produces quark masses of order
$\Lambda_{QCD}$ or less [for a magnetic field all the way up to the order
of $(10 \mbox{ TeV})^2$]. The point is that these masses are heavy in
units
of the new confinement scale $\lambda_{QCD}$ and the pairing dynamics
is indeed weakly coupled.

\section{QCD with large number of colors}
\label{infty}

In this section, we will discuss the dynamics in QCD in a magnetic field
when the number of colors is large, in particular, we will consider the
('t Hooft) limit $N_c \to \infty$. Just a look at expression (\ref{M_g})
for the gluon mass is enough to recognize that the dynamics in this limit
is very different from that considered in the previous sections. Indeed,
as is well known, the strong coupling constant $\alpha_s$ is proportional
to $1/N_c$ in this limit. More precisely, it rescales as
\be
\alpha_s = \frac{\tilde{\alpha}_s}{N_c},
\label{tilde}
\ee
where the new coupling constant $\tilde{\alpha}_s$ remains finite
as $N_c \to \infty$. Then, expression (\ref{M_g}) implies that
the gluon mass goes to zero in this limit. This in turn implies 
that the appropriate approximation in this limit is not
the improved rainbow approximation but the rainbow approximation
itself, when {\it both} the vertex and the gluon propagator in
the SD equation (\ref{SD}) are taken to be bare.

In order to get the expression for the quark in this case, 
we can use the results of the analysis of the SD equation in the 
rainbow approximation in QED in a magnetic field
\cite{QED1}, with the same simple modifications as in
Sec. \ref{mag-cat}. The result is
\be
m_{q}^2 = C |e_{q}B|
\exp\left[-{\pi}
\left(\frac{\pi N_c}{(N_{c}^2-1)\alpha_s}\right)^{1/2}
\right],
\label{m_q}
\ee
where the constant $C$ is of order one. As $N_c \to \infty$, one
gets
\be
m_{q,\infty}^2 = C |e_{q}B|
\exp\left[-{\pi}
\left(\frac{\pi}{\tilde{\alpha}_s}\right)^{1/2}
\right].
\label{m{infty}}
\ee
It is natural to ask how large $N_c$ should be before the expression 
(\ref{m_q}) becomes reliable. From 
our discussion above, it is clear that the rainbow 
approximation may be reliable only when the gluon mass is small, i.e.,
it is of the order of the quark mass $m_q$ or less.
Equating expressions (\ref{M_g}) and (\ref{m_q}), we derive an
estimate for the threshold value of $N_c$:
\be
N_{c}^{thr} \sim \frac{2N_u + N_d}{\ln|eB|/\Lambda^{2}_{QCD}} 
\exp\left[\frac{\pi}{2\sqrt{3}}
\left(11\ln\frac{|eB|}{\Lambda_{QCD}}\right)^{1/2}
\right].
\label{estimate1}
\ee
Expression (\ref{m_q}) for the quark mass
is reliable for the values of 
$N_c$ of the order of $N_{c}^{thr}$ or larger.
Decreasing $N_c$ below $N_{c}^{thr}$, one comes to
expression (\ref{gap}). 

It is quite remarkable that one can get a rather
close estimate for $N_{c}^{thr}$ by equating expressions
(\ref{gap}) and (\ref{M_g}):
\be
N_{c}^{thr} \sim \frac{2N_u + N_d}{\ln|eB|/\Lambda^{2}_{QCD}}
\exp\left[\left(11\ln\frac{|eB|}{\Lambda_{QCD}}\right)^{1/2}\right]
\label{estimate2}
\ee
[notice that the ratio of the exponents (\ref{estimate1})
and (\ref{estimate2}) is equal to $0.91$].
The similarity of estimates (\ref{estimate1}) and (\ref{estimate2})  
implies that, crossing the threshold $N_{c}^{thr}$, expression
(\ref{m_q}) for $m_q$ smoothly transfers into expression (\ref{gap}).

These estimates show that the value of $N_{c}^{thr}$ rapidly grows with
the magnetic field. For example, taking $\Lambda_{QCD} = 250 \mbox{ MeV}$
and $N_{u}=1$, $N_{d}=2$, we find from Eq. (\ref{estimate1}) that $N^{thr}_c
\sim 10^2$, $10^3$, and $10^4$ for $|eB| \sim (1\mbox{ GeV})^2$, $(10\mbox{
GeV})^2$, and $(100\mbox{ GeV})^2$, respectively.

As was shown in Sec. \ref{conf}, in the regime with the number
of colors $N_c \ll N_{c}^{thr}$, the confinement scale $\lambda_{QCD}$ in
QCD in a strong magnetic field is essentially smaller than
$\Lambda_{QCD}$. What is the value of $\lambda_{QCD}$ in the regime with
$N_c$ being of the order of $N_{c}^{thr}$ or larger? It is not difficult
to see that $\lambda_{QCD} \simeq \Lambda_{QCD}$ in this case. Indeed, now
the gluon mass and, therefore, the contribution of quarks in the
polarization operator are small (the latter is suppressed by the factor
$1/N_c$ with respect to the contribution of gluons). As a result, the
$\beta$-function in this theory is close to that in QCD without a magnetic
field, i.e., $\lambda_{QCD} \simeq \Lambda_{QCD}$.

Expression (\ref{m_q}) implies that, for a sufficiently strong
magnetic fields, the dynamical mass $m_q$ is much larger than
the confinement scale $\Lambda_{QCD}$. Indeed, expressing the 
magnetic field in terms of the running coupling, one gets
\be
m_{q}^2 \simeq \left|\frac{e_{q}}{e}\right| \Lambda_{QCD}^2
\exp\left[\frac{1}{b\alpha_{s}} -{\pi}
\left(\frac{\pi N_c}{(N_{c}^2-1)\alpha_s}\right)^{1/2}
\right],
\ee
and for small values of $b\alpha_s \sim N_{c}\alpha_{s} \equiv
\tilde{\alpha}_s$ (i.e., for large values of $|eB|$) the mass $m_q$ is
indeed much larger than $\Lambda_{QCD}$. This point is important for
proving the reliability of the rainbow approximation in this problem.
Indeed, the relevant region of momenta in this problem is $m_{q}^2 \ll
|k^2| \ll |eB|$ \cite{QED1} where, 
because $m_{q}^2 \gg \Lambda_{QCD}^2$ for a strong enough field, 
the running coupling is small. Therefore the rainbow approximation is
indeed reliable for sufficiently strong magnetic fields in
this case.

\section{Conclusion}

QCD in a strong magnetic field yields an example of a rich, sophisticated
and (that is very important) controllable dynamics. Because of the
property of asymptotic freedom, the pairing dynamics, responsible for
chiral symmetry breaking in a strong magnetic field, is weakly
interacting. The key point why this weakly interacting dynamics manages to
produce spontaneous chiral symmetry breaking is the fact that it is
essentially 1+1 dimensional: in the plane orthogonal to the external field
the motion of charged quarks is restricted to a region of a typical size
of the order of the magnetic length, $\ell = 1/\sqrt{|e_{q}B|}$. Moreover,
such a dynamics almost completely decouples from the dynamics of
confinement which develops at very low energy scales in the presence of a
strong magnetic field.

While the pairing dynamics decouples from the dynamics of confinement, the
latter is strongly modified by the polarization effects due to quarks that
have an effective $1+1$ dimensional dynamics in the lowest Landau level.
As a result, the confinement scale in QCD in a strong magnetic field
$\lambda_{QCD}$ is much less than the confinement scale $\Lambda_{QCD}$ in
the vacuum QCD. This implies a rich spectrum of light glueballs in this
theory.

This picture changes drastically for QCD with a large number of colors
($N_c \gtrsim 100$ for $|eB| \gtrsim 1 \mbox{ GeV}^2$). In that case a
conventional confinement dynamics, with the confinement scale 
$\lambda_{QCD} \sim \Lambda_{QCD}$, is realized.
 
The dynamics of chiral symmetry breaking in QCD in a magnetic field has
both similarities and important differencies with respect to the dynamics
of color superconductivity in QCD with a large baryon density \cite{RW}.
Both dynamics are essentially 1+1 dimensional. However, while the former
is anisotropic [the rotational $SO(3)$ symmetry is explicitly broken by a
magnetic field], the rotational symmetry is preserved in the latter.
This fact is in particular connected with that while in dense QCD
quarks interact both with chromo-electric and chromo-magnetic
gluons \cite{hard}, in the present theory they interact only with the 
longitudinal components of chromo-electric gluons. This in turn
leads to very different expressions for the dynamical masses of
quarks in these two theories.   

Another important difference is that while the pseudo-Higgs effect takes 
place in QCD in a magnetic field, the genuine Higgs (Meissner-Higgs) effect
is realized in color superconducting dense quark matter. Because of the 
Higgs effect, the color interactions connected with broken generators are
completely screened in infrared in the case of color superconductivity. 
In particular, in the color-flavor locked
phase of dense QCD with three quark flavors, the color symmetry is
completely broken and, therefore, the infrared dynamics is under control
in that case \cite{ARW}. As for dense QCD with two quark
flavors, the color symmetry is only partialy broken down to $SU(2)_c$, 
and there exists an analog of the pseudo-Higgs effect for the electric
modes of gluons connected with the unbroken $SU(2)_c$. As a result, the
confinement scale of the gluodynamics of the remaining $SU(2)_c$ group is
much less than $\Lambda_{QCD}$ \cite{RSS}, like in the present case. The
essential difference, however, is that, unlike QCD in a magnetic field,
the infrared dynamics of a color superconductor is isotropic.

Last but not least, unlike the case of QCD with a nonzero baryon density,
there are no principal obstacles for examining all these results and
predictions in lattice computer simulations of QCD in a magnetic field.

\begin{acknowledgments}
We thank Arkady Vainshtein for raising a question concerning the
dynamics in QCD with $N_{c} \to \infty$ in a magnetic field. V.A.M. is
grateful for support from the Natural Sciences and Engineering Research
Council of Canada. The work of I.A.S. was supported by the U.S. Department
of Energy under Grant No.~DE-FG02-87ER40328 and in part by the National
Science Foundation under Grant No.~PHY99-07949. 
\end{acknowledgments}


\begin{references}
\bibitem{footnote} The fact that the pairing dynamics responsible for
generating rather light quark masses (of
order $\Lambda_{QCD}$ or less) decouples from the confinement
dynamics may seem to be
puzzling. As it will be shown in Sec. \ref{conf}, the confinement
scale $\lambda_{QCD}$ in QCD in a strong magnetic field is
much less than $\Lambda_{QCD}$, i.e., the quark masses are
large if expressed in units of the new confinement scale. This
point implies that the pairing dynamics is indeed
weakly coupled.
\end{references}
\end{document}